\documentclass[11pt,twoside]{article}
\usepackage{asp2006}
\usepackage{epsf}
\usepackage{lscape}

\begin{document}

\title{Search For Iron, Nickel, and Fluorine in PG1159 Stars}   

\author{E. Reiff,$^1$ T. Rauch,$^1$ K. Werner,$^1$ J.~W. Kruk,$^2$ and L. Koesterke$^3$}  
\affil{$^1$Institute for Astronomy and Astrophysics, Kepler Center for
  Astro and Particle Physics, University of T\"ubingen, Sand~1, 72076 T\"ubingen, Germany\\
$^2$Department of Physics and Astronomy, Johns
Hopkins University, Baltimore, MD21218, U.S.A.\\
$^3$McDonald Observatory and Department of 
Astronomy, University of Texas, Austin, TX78712-1083, U.S.A.
}    

\begin{abstract} 
A possible origin of the iron-deficiency in PG1159 stars could be
neutron captures on Fe nuclei. A nickel overabundance would corroborate
this idea. Consequently we are looking for nickel lines in PG1159
stars. Prime targets are relatively cool objects, because Ni\,{\sc vi}
is the dominant ionisation stage and the spectral lines of this ion are
accessible with UV observations. We do not find such lines in the coolest
PG1159 star observed by FUSE (PG1707+427, $T_{\rm eff} = 85\,000$~K) and
conclude that the nickel abundance is not enhanced. Hence, the
Fe-deficiency in PG1159 stars remains unexplained.
In addition, we present results of a wind analysis of the hybrid-PG1159
star NGC\,7094 and the [WC]--PG1159 transition-type object Abell\,78 in
order to derive F abundances from the F\,{\sc vi} 1139.5\,\AA\
line. In both cases, we find F overabundances, in agreement with
results of photospheric analyses of many PG1159 stars. Surprisingly, we
find indications for a very low O abundance in NGC~7094. 
\end{abstract}

\section{Search for iron and nickel in ``cool'' PG1159 stars} 

It is widely believed
that PG1159 stars exhibit former AGB-star intershell matter on their surface,
dredged-up by a late He-shell flash.  One of the most
surprising results of spectral analyses of PG1159 stars is the observed
iron-deficiency in all objects investigated so far. In not a single case
were Fe lines identified. The derived upper limits suggest
Fe-deficiencies of the order 0.5--1.5~dex. (For a more detailed
description see the PG1159 review in
these proceedings by Werner et al.) One possible explanation is that
Fe was transformed to heavier elements by s-process neutron captures
during the preceding AGB evolution. However, AGB star evolution
models do not predict such a strong Fe depletion in the
intershell. Therefore, either our understanding of the s-process in AGB
stars has a fundamental flaw, or the Fe-deficiency has another
origin. One hint to the solution of this problem is the
nickel abundance in PG1159 stars. If it is oversolar, then the
hypothesis of n-captures on Fe seed nuclei would be verified. If we do not find a Ni enhancement, this hypothesis
would not necessarily be disproved, because Fe might have been transformed
into even heavier elements.

In contrast to the iron-abundance analysis, the nickel abundance
determination in PG1159 stars is more difficult. The iron-deficiency was
derived from the absence of Fe\,{\sc vii} lines in UV spectra (obtained
with HST and FUSE), because  Fe\,{\sc vii} is the dominant ionisation
stage. From model predictions  Ni\,{\sc vii} lines should be the
strongest signatures of nickel but, unfortunately, these lines are
located at wavelengths below the Lyman edge and, thus, are not
accessible. An alternative is offered by the coolest PG1159 stars, for
which Ni\,{\sc vi} lines are the dominant features, which are located in
the FUSE spectral range. We have therefore investigated the spectrum of
PG1707+427, which is the coolest PG1159 star observed by FUSE ($T_{\rm
eff} = 85\,000$~K, $\log g=7.5$). We also analysed another relatively cool PG1159
star, PG1424+535, which is significantly hotter ($T_{\rm eff} =
110\,000$~K, $\log g=7.5$) and, hence, less promising for our purpose.

\begin{figure}[t]
\plottwo{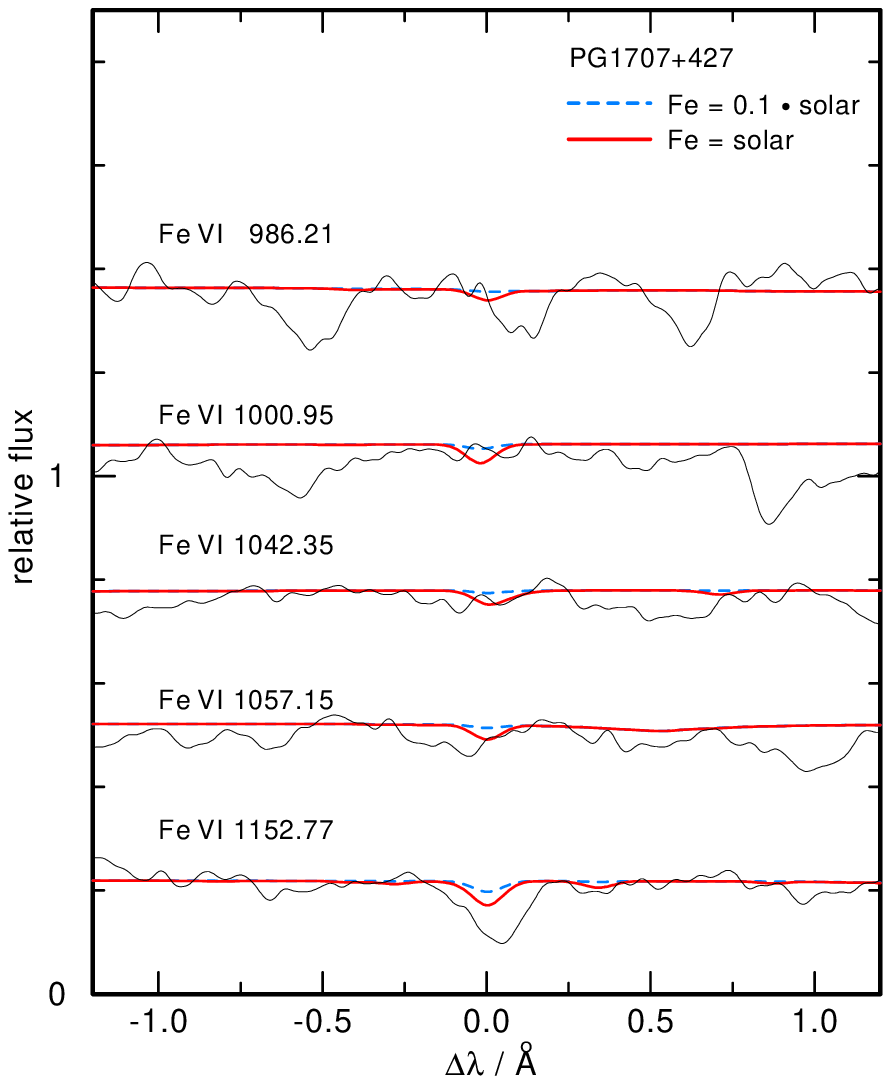}{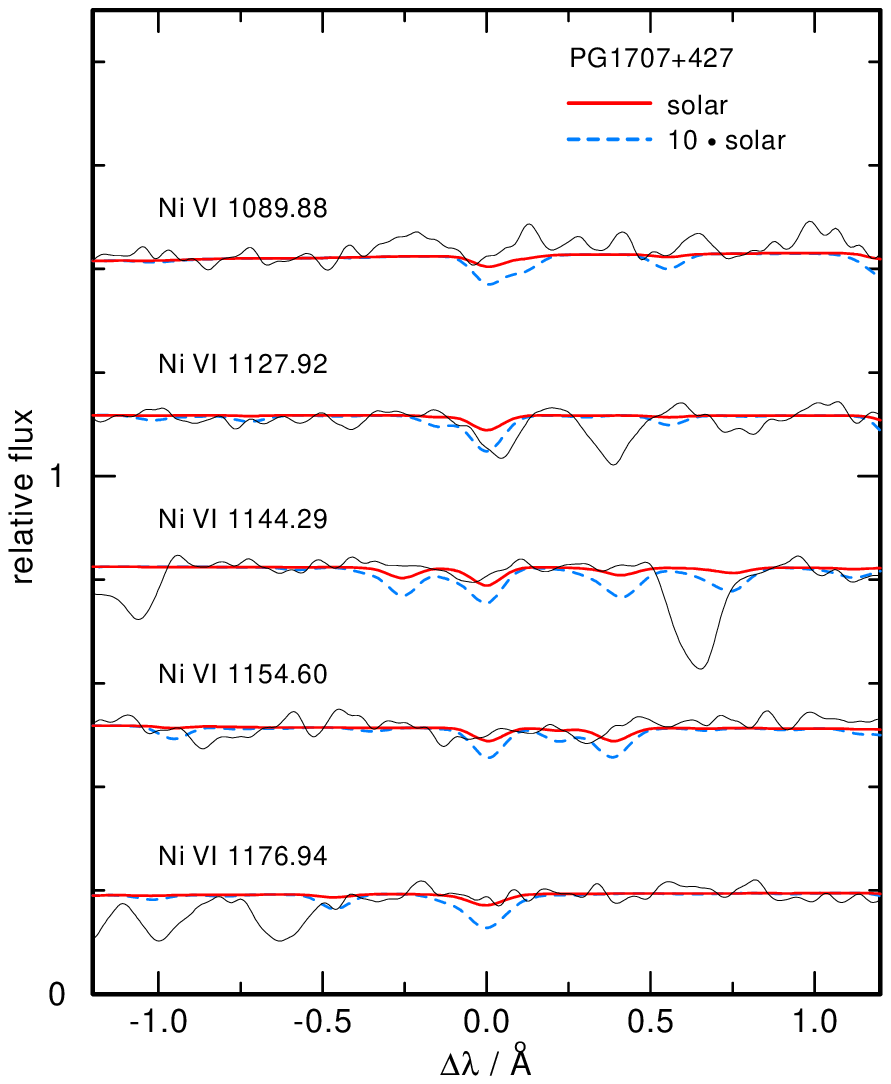}
\caption[]{Sections of the FUSE spectrum of PG1707+427, centered on
  the strongest expected iron lines (left panel) and nickel lines (right
  panel). Neither species is identified. Overplotted
  are models with Fe and Ni abundances as given in the panels. While
  iron appears to be depleted, there is no simultaneous nickel
  enhancement.  Other element abundances in the model are He/C/N/O/Ne
  = 0.42/0.38/0.015/0.17/0.02 (mass fractions).}      
\label{fig:nickel} 
\end{figure}

Our model atmospheres are plane-parallel, homogeneous non-LTE models, in
hydrostatic and radiative equilibrium (Werner et
al. 2003). The composition comprises the main atmospheric constituents
He, C, N, O and Ne. The atmospheric parameters were taken from previous
analyses (Werner \& Herwig 2006).  Fe and Ni were included following
Rauch \& Deetjen (2003) and keeping
fixed the atmospheric structure. 

Fig. \ref{fig:nickel} shows the
comparison of PG1707+427 to models with different Fe and Ni
abundances. Lines from neither species can be identified. The derived
upper limit for both elements is the solar value. Nickel is
certainly not oversolar, however, we cannot prove that this particular
star is Fe-deficient. If it were, then the result on nickel would not
corroborate the n-capture hypothesis for the Fe-deficiency. In conclusion, our
results do not answer the question why PG1159 stars are generally
Fe-deficient. In the case of PG1424+535, the absence of  Fe\,{\sc vii}
lines means an Fe-deficiency of 1~dex but, for the reasons mentioned
above, no meaningful limit on the Ni abundance can be set.

\begin{table}[t]
\caption{Parameters of the wind models for Abell\,78 and
  NGC\,7094 (Koesterke et al.\@ 1998). F abundances are derived in this work. ($\log$\,F$_\odot$=-6.3)}
\label{tab:windpara}
\footnotesize
\begin{center}
\begin{tabular}{lccccccc}
\noalign{\smallskip}
\tableline
\noalign{\smallskip}
Object & Spectral &\multicolumn{1}{c}{$ T_{\rm eff} $}
       & \multicolumn{1}{c}{$ \log g $} 
       & \multicolumn{1}{c}{$ \log L $} 
       & \multicolumn{1}{c}{$ \log {\dot M} $} 
       & \multicolumn{1}{c}{$ v_{\infty} $}  
       & \multicolumn{1}{c}{$ \log $ F}  \\ 

       & Type & \multicolumn{1}{c}{ [kK] }
       & \multicolumn{1}{c}{ [cm/s$^2$]} 
       & \multicolumn{1}{c}{ [L$_{\odot}] $} 
       & \multicolumn{1}{c}{ [M$_{\odot}$/yr]} 
       & \multicolumn{1}{c}{ [km/s]}  
       & \multicolumn{1}{c}{[mass fraction]}  \\ 
\noalign{\smallskip}
\tableline
\noalign{\smallskip}
Abell 78  & [WC]--PG1159 & 110 & 5.5 & 3.7 & -7.3 & 3750 & -5.9\\ 
NGC\,7094 & hybrid PG1159& 110 & 5.7 & 3.6 & -7.7 & 3500 & -5.1\\ 
\noalign{\smallskip}
\tableline
\end{tabular}
\end{center}
\end{table}

\begin{figure}[t]
\begin{center}
\epsfxsize=9cm \epsffile{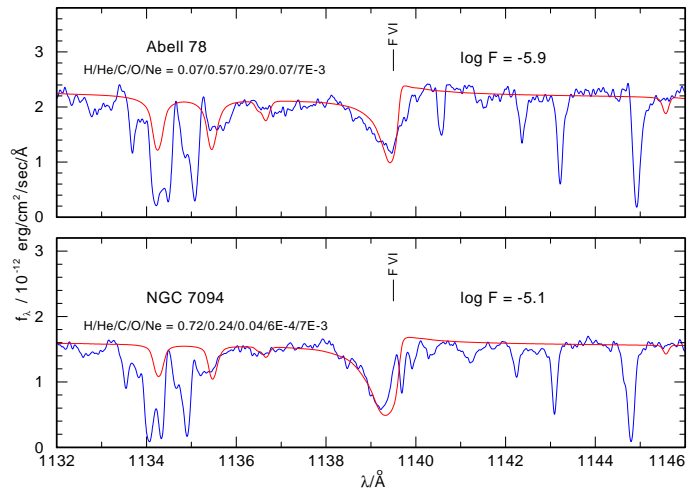}
\end{center}
\vspace{-0.5cm}
\caption[]{Details from FUSE spectra of the [WC]--PG1159 transition-type
  object Abell~78 (top panel) and the hybrid PG1159 star NGC\,7094
  (bottom) near the  F\,{\sc vi} 1139.5\,\AA\ line. Overplotted are synthetic spectra with model
  parameters as given in Tab.~\ref{tab:windpara}. 
}
\label{fig:windf6} 
\end{figure}

\section{Wind models for NGC\,7094 and Abell~78} 

Several PG1159 objects display strong P Cygni wind profiles in their
spectra, among them the central stars of Abell\,78 and NGC\,7094. Their parameters are summarized in
Tab.\,\ref{tab:windpara}. Based on these values we calculated expanding
stellar atmosphere models using the ``Hot Blast'' wind code of Koesterke.

Abell\,78 is a
transition-type object which can be considered as a [WC] star in the
phase of becoming a PG1159 star. The abundance of the main atmospheric
constituents confirms this idea. The discovery that fluorine in many
PG1159 stars is strongly overabundant is interesting in the context of
AGB-star nucleosynthesis (Werner et al.\@ 2005). The respective analyses
were based on a newly discovered F\,{\sc vi} line at 1139.5~\AA\ with
plane-parallel atmosphere models. The line was also discovered in
Abell~78, however, the asymmetric shape suggests that the
line is formed in the stellar wind. The top panel of
Fig.\,\ref{fig:windf6} displays the FUSE spectrum together with a model
fit. F is 0.4~dex oversolar.

\begin{figure}[t]
\begin{center}
\epsfxsize=11cm \epsffile{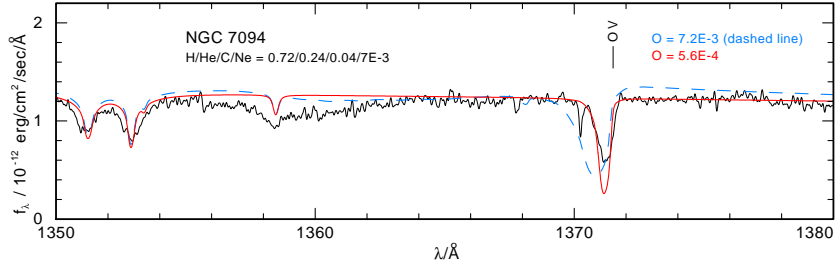}
\end{center}
\vspace{-0.5cm}
\caption[]{The O\,{\sc v} 1371\,\AA\ line in the HST/STIS
  spectrum of NGC\,7094. Overplotted are two models with different
  O abundance as depicted (mass fractions). For other model parameters see
  Tab. \ref{tab:windpara}. }     
\label{fig:windo5} 
\end{figure}

NGC\,7094 is a hybrid-PG1159 star, denoting
objects in the PG1159 spectral class which have a detectable amount of
residual hydrogen in their atmospheres. In contrast to the other PG1159
stars, which are the result of a late or very late thermal pulse (LTP
and VLTP), hybrid-PG1159 stars are thought to be the outcome of a final
TP occurring on the AGB (AFTP) immediately before the star's departure of
the AGB. Quite unexpectedly (because of residual H-envelope material
which should also contain Fe in a solar Fe/H ratio), Abell~78 is
strongly Fe-deficient like other PG1159 stars (see Ziegler et al.\@ in
these proceedings). NGC\,7094 also shows a strong F\,{\sc vi} 1139.5\,\AA\
line, and we derive a 1.2 dex oversolar value. Like the Fe-deficiency,
the strong F enrichment is difficult to understand in the AFTP picture.

On the other hand, the low oxygen abundance in NGC~7094 as compared to
other PG1159 stars corroborates the AFTP scenario. Dreizler et al.\
(1995) found O$\approx$0.01 (mass fraction) from optical
spectroscopy. We have computed wind models with different O abundances
in order to fit the O\,{\sc v} 1371\,\AA\ line
(Fig.\,\ref{fig:windo5}). We conclude that the O abundance is even lower
than hitherto thought, by about 1 dex. This means that it is even
\emph{below} the solar value ($5.3\cdot10^{-3}$, Asplund et al.\@ 2005) so that, in contrast to
PG1159 stars, O is not increased by dredge-up processes from the C-O
stellar core. However, this result must be regarded as preliminary,
because other oxygen lines still need to be investigated.

\acknowledgements{T.R.\@ is supported by the {\it German Astrophysical
    Virtual Observatory} (BMBF grant 05 AC6VTB), E.R.\@
    by DFG grant We1312/30$-$1, J.W.K.\@ by the
    FUSE project, funded by NASA contract NAS5-32985.}

\end{document}